This paper is aimed at working out a simple though reliable vibronic model appropriate to describing the dipolar relaxation in alkali halides by means of the reaction rate theory. Assuming coupling to a single promoting mode, the Hamiltonian of an isolated dipole in a crystal is simplified so as to render the problem 1-D and easily solvable. The electron-mode interaction is accounted for by up to 2nd-order terms of the expansion in the mode coordinate so as to allow for changes of the dressed-oscillator vibrational frequency, as the electronic state of the dipole is changed on photoexcitation. The crossover splitting $V_{12}$ controlling the reorientation is found to be twice the modulus of the off-diagonal matrix element of the interaction Hamiltonian. Depending on $V_{12}$, dipoles may reorientate either adiabatically with a high electron-transfer expectancy, or may exhibit low reorientation rates due to non-adiabaticity because of low electron-transfer probabilities. An important quantity to distinguish between adiabatic dipoles which reorientate classically by over barrier jumps and ones which reorientate by configurational tunneling is Christov's temperature $T_c$ at which the classical rate is just equal to the tunneling rate. We believe to have arrived at an useful formula relating $T_c$ to the barrier height $E_b$ and the saddle-point splitting $V_{12}$ which enables one to extract virtually all the reorientational parameters from experimental pre-exponential factors and activation energies. To illustrate the capacity of the method, ITC data on the relaxation of impurity-vacancy (I-V) dipoles in $Eu^{2+}$-doped alkali halides obtained at UNAM are reanalyzed. The dipoles display an increasing trend to improve adiabaticity and behave less classically, as the reorientational energies increase. The temperature dependences predicted by theory agree with ones obtained by experiment.

1. Introduction

Low-symmetry dipolar defects in alkali halides have been the subject of considerable experimental and theoretical interest for some time [1-3]. These species produce an anisotropic lattice distortion around themselves and, in some cases, carry an electric dipole by virtue of which they can reorientate in an applied external stress or electric field. Under certain conditions (temperature, field strength) considerable degrees of dipole alignment can be achieved and experimentally observed. The techniques employed have mainly made use of the (i) dielectric loss, (ii) ionic thermocurrents (ITC), (iii) electrocaloric effect, and (iv) field- or (v) light-induced dichroism in the optical absorption spectra [4,5]. The reorientational motion of a dipole occurs between more- or less- well defined discrete angular positions in the crystalline lattice. From a classical point of view, the transition of a localized dipole from one such position to the next materializes via classical jumps or by quantum-mechanical tunneling across some potential-energy barrier peaked in-between. This transfer interaction lifts the degeneracy of the quantum states of the separate wells, formed between subsequent barriers along the angular coordinate, by inducing a splitting of the energy levels, the resulting mixed-up states already spreading over all the wells. Now, the common method of the above-mentioned techniques is to create a population difference by disrupting the equilibrium distribution of the occupation probabilities of the energy levels, as the external perturbation is switched-on, and then follow the recovery of thermal equilibrium, as that perturbation is switched-off. The relaxation time $\tau_{rel}$ derived from the experimental data more or less directly provides virtually all the desired information on the transfer rates and level splittings. In particular, its temperature dependence tells a lot about the nature of both the reorientation-promoting mode and its interaction with the rotating entity. At the same time, the relaxation-time data provide information on the symmetry of the dipole, that is, of its reorientational sites, when measured using (iii) through (v) [4,5]. Based on the overall data collected in this way, some dipoles have been proclaimed "classical", their relaxation times being found to depend on the temperature in the Arrhenius manner, while others have proved to rapidly reorientate by quantum-mechanical tunneling at low temperatures. Because of the inherent requirement that the dipolar system remains in a frozen-in polarized state for sufficiently long a time before an ITC run, the technique under (ii) is usually considered nonapplicable to the study of "quantal dipoles".

Among the variety of dipolar defects whose reorientational motion has so far been investigated experimentally one can extract four main groups, as follows: (a) I-V dipoles formed by an aliovalent impurity ion and the compensating vacancy in either sublattice; (b) substitutional molecular ions with an intrinsic electric dipole momemt; (c) electric dipoles formed when an unvalent atomic impurity ion occupies an off-center position; (d) $F_A$-type color centers [4-6].

The common theoretical treatment of the dipolar relaxation problem has made use of the multiphonon approach based on perturbation theory [7]. Lately, however, a new method has been proposed [8,9] employing the reaction rate theory [10]. The reaction-rate approach (RRA) is simpler and helps to easily reveal the physics. By making no use of perturbation methods, RRA is inherently applicable to both adiabatic (strong perturbation) and nonadiabatic (weak perturbation) transitions, as well as to the variety in-between.

The present method is a continuation of our earlier work [9], and is mainly aimed at reassessing the common attitude to "classical dipoles" from the RRA viewpoint in the quest for possible "hidden" quantal or adiabaticity effects. In this respect, dipolar systems will be pointed out where such effects can possibly occur and be observed in an explicit form by means of appropriately planned experiments. Some considerations will also be presented regarding the photoinduced reorientation in which the dipolar relaxation occurs in the excited electronic state of the component defect. These will concentrate on a qualitative assessment of the reorientational barriers and rotational frequencies. While most of our discussion will potentially apply to dipoles studied by all the five techniques mentioned under (i) through (v), the emphasis will be laid on the ITC method which, because of its inherent sensitivity, has so far been the main data-collecting tool [11].

## 2. Hamiltonian background

We consider a dipole embedded in a crystalline medium regarded as a system of lattice oscillators (vibrating ions) plus outer-shell electrons, coupled to the oscillators. The relevant Hamiltonian will contain electron, lattice, and electron-lattice terms, respectively,

$$H = H_e + H_L + H_{eL} \qquad (1)$$

$H_e$ is the static electronic Hamiltonian at a fixed lattice when all the oscillators are frozen-in at their unperturbed equilibrium positions at $\tilde{q} = 0$:

$$H_e = \Sigma[\mathbf{p}_e^2/2m_e + V_e(\mathbf{r}_e, \tilde{0})] \qquad (2)$$

the sum being over the coordinates $\mathbf{r}_e$ and momenta $\mathbf{p}_e$ of all the electrons; $m_e$ are the electron masses. $V_e(\mathbf{r}_e, \tilde{0})$ is the static potential which the electrons "see" when the nuclei are at rest. When they are not, the electronic potential varies following parametrically the motion of the nuclei (adiabatic approximation). This modulated potential $V_e(\mathbf{r}_e, \tilde{q})$ can be expanded into a power series in $\tilde{q}$ (the domain of nuclear coordinates) to give

$$V_e(\mathbf{r}_e, \tilde{q}) = V_e(\mathbf{r}_e, \tilde{0}) + \tilde{b}(\mathbf{r}_e).\tilde{q} + \tilde{q}.\tilde{c}(\mathbf{r}_e).\tilde{q} + ... \qquad (3)$$

in terms of vector-vector and tensor-vector products, etc. The mixed electron-lattice terms in (3) effect the coupling of the electrons to the lattice oscillators. Consequently,

$$H_{eL} = \Sigma[\tilde{b}(\mathbf{r}_e).\tilde{q} + \tilde{q}.c(\mathbf{r}_e).\tilde{q} + ...] \qquad (4)$$

The lattice Hamiltonian in the absence of an electron-lattice interaction is

$$H_L = \Sigma_s[\mathbf{P}_s^2/2M_s + \tfrac{1}{2}M_s\omega_s^2 q_s^2] + \mathbf{P}^2/2M + \tfrac{1}{2}M\omega^2 q^2 + ... \qquad (5)$$

where the reorientation-promoting mode is taken out of the sum (subscripts omitted). Here $q_s$, $\mathbf{P}_s$, $\omega_s$, and $M_s$ stand for the coordinates, momenta, angular frequencies, and masses of the 'bare' lattice oscillators, respectively. We further make several assumptions to simplify the mathematical problem without sacrificing physics:

(i) Harmonic approximation: omitting the dots in (5), while considering $H_L$ to be diagonalized with respect to the nuclear coordinates.

(ii) Predominance of the electron-phonon interaction for the promoting mode.

(iii) Adopting a coupling scheme that confines to considering linear and second-order terms only, thus neglecting the dots in (3).

In the absence of a lattice-lattice interaction under (i), (ii) makes considering the sum-terms in (5) unimportant, and (1) reduces to

$$H = H_e + b(\mathbf{r})q + c(\mathbf{r})q^2 + \mathbf{P}^2/2M + \tfrac{1}{2}M\omega^2 q^2 \qquad (6)$$

This simplifies the relaxational problem to the extent of making it 1-D (one-dimensional).

To solve the (6)-based Schrodinger equation within the Born-Oppenheimer approximation we first define an 'adiabatic Hamiltonian' by means of

$$H_{ad} = H - T_L = H_e + b(\mathbf{r})q + c(\mathbf{r})q^2 + \tfrac{1}{2}M\omega^2 q^2 \qquad (7)$$

where $T_L$ is the nuclear kinetic energy. Following a well-known prescription [12], we obtain the 'diabatic surface' for an electronic state j:

$$V_j(q) = <j,0 | H_{ad} | j,0> = E_j + b_j q + c_j q^2 + \tfrac{1}{2} M\omega^2 q^2 \qquad (8)$$

where

$$E_j = <j,0 | H_e | j,0> \qquad (9)$$

is the eigenvalue in eigenstate $|j,0>$ of the static Hamiltonian (2). The $b_j$'s and $c_j$'s are, correspondingly, the first-order (linear)

$$b_j = <j,0 | b(r) | j,0> \qquad (10)$$

and second-order (quadratic)

$$c_j = <j,0 | c(r) | j,0> \qquad (11)$$

electron-phonon coupling coefficients. Equation (8) is that of a second-degree parabola, whose minimum is displaced from the origin (q = 0) at

$$q_j = -b_j / M\omega_j 2 \qquad (12)$$

This is the new equilibrium position of the lattice oscillator which displaces from the origin as a result of the electron-phonon interaction. The angular frequency of the electron-coupled oscillator also changes from the 'bare' value to

$$\omega_j = (\omega^2 + 2c_j / M)^{1/2} \qquad (13)$$

While the latter is a second-order effect, the displacement of the equilibrium coordinate results essentially from the linear coupling. Using (12) and (13), equation (8) becomes

$$V_j(q) = \tfrac{1}{2} M\omega_j^2 (q - q_j)^2 + Q_j \qquad (14)$$

where

$$Q_j = E_j - \tfrac{1}{2} M\omega_j^2 q_j^2 = E_j - \tfrac{1}{2} (M\omega_j^2)^{-1} b_j^2 \qquad (15)$$

is the electron-binding energy within the adiabatic approximation.

To define two neighboring reorientational sites 1 and 2 we need two static electronic states $j_1$ and $j_2$ and set $Q_1 = Q_2$ (subscript 'j' omitted). Ultimately, this will bring about a symmetric double-well situation which, however, may have to be altered if more than one hopping mechanisms are involved: in this latter case the reaction heat $Q = Q_1 - Q_2$ may have to be taken into account. For the sake of simplicity we take up the symmetric case. From (15) we get, assuming the same 'dressed' phonon frequency $\omega_j$ in both 1 and 2:

$$E_2 - E_1 = \tfrac{1}{2} M\omega_j^2 (q_2^2 - q_1^2) = M\omega_j^2 \Delta q q_c \qquad (16)$$

where $q_c = \frac{1}{2}(q_1+q_2)$ is the crossover lattice configuration, while $\Delta q = (q_2-q_1)$ is the separation between the minima of $V_1(q)$ and $V_2(q)$ along the relaxation coordinate. The corresponding equations for $V_1(q)$ and $V_2(q)$ obtain from (14) at $j_1$ and $j_2$. Eventually, the static states $j_1$ and $j_2$ split from j self-consistently because of the tunneling electron-exchange interaction.

On studying the relaxational behavior of the dipolar system, well 1 can initially be over populated relative to well 2, the population difference being created through an external perturbation following the methods discussed in Section 1. This will give rise to an effective flow of vibrons (dipoles) from 1 to 2 until thermal equilibrium is restored and the population difference nullified. The surplus energy is given away to the lattice during relaxation by virtue of the promoting mode - accepting modes interactions omitted on cancelling the dots in (3). These lead to intrawell vertical transitions, which, even though not explicitly taken into account, are always implied if fast enough so as to make the horizontal vibronic transitions between 1 and 2 the rate-determining step. However, the latter interwell transitions could not materialize unless there is a finite electron coupling between the eigenstates $j_1$ and $j_2$ of the static Hamiltonian (2) in the sense that the electron-exchange matrix element

$$V_{12}(q) = <j_2,0 \mid H_{ad} \mid j_1,0> \tag{17}$$

is nonvanishing at least in the 'transition region' around the crossover configuration $q_c$ between $V_1(q)$ and $V_2(q)$.

Now, using

$$\mid j,q> = A_1(q) \mid j_1,0> + A_2(q) \mid j_2,0> \tag{18}$$

to solve Schrodinger's equation

$$H_{ad} \mid j,q> = E \mid j,q> \tag{19}$$

we obtain following the usual procedure

$$[V_{11}(q) - E]A_1(q) + V_{21}(q)A_2(q) = 0$$

$$V_{12}(q)A_1(q) + [V_{22}(q) - E]A_2(q) = 0$$

where orthonormality of $\mid j_1,0>$ and $\mid j_2,0>$ has been presumed. Here $V_{11}(q)$ stands for $V_1(q)$, etc. The resulting secular equation is solved to give two adiabatic surfaces, lower

$$E_L(q) = \frac{1}{2}\{V_{11}(q)+V_{22}(q) - ([V_{11}(q) - V_{22}(q)]^2 + 4|V_{12}(q)|^2)^{1/2}\} \tag{20}$$

and upper

$$E_L(q) = \frac{1}{2}\{V_{11}(q)+V_{22}(q) + ([V_{11}(q) - V_{22}(q)]^2 + 4|V_{12}(q)|^2)^{1/2}\} \tag{21}$$

From (17) and (7) we get

$V_{12}(q) = <j_2 | H_{eL} | j_1> = b_{12}q + c_{12}q^2 + ...$ (22)

where

$b_{12} = <j_2 | b(\mathbf{r}) | j_1>$ (23)

and

$c_{12} = <j_2 | c(\mathbf{r}) | j_1>$ (24)

etc. are the off-diagonal coupling constants. They will be assumed small, $c_{12}$ virtually vanishing. From (14) we have

$V_1(q) - V_2(q) = M\omega_j^2(q_2 - q_1)q - (E_2 - E_1)$ (25)

so that the 'small $b_{12}$' condition reads

$|b_{12}| << \frac{1}{2} M\omega_j^2 | q_2 - q_1 | = \frac{1}{2} | b_2 - b_1|$ (26)

Under (26) $E_L(q)$ will coincide with the corresponding diabatic branches, except for a certain 'transition region' around the crossover ($q_c$). The gap between $E_U(q)$ and $E_L(q)$ amounts to $2 | V_{12}(q_c) |$ at crossover. A barrier of energy

$E_c = E_c' - |V_{12}| = (2M\omega_j^2)^{-1} \{\frac{1}{4}(b_2 - 3b_1) - | b_{12} |sgn(b_1+b_2)\}(b_1+b_2) + E_1$ (27)

forms on the lower adiabatic surface peaked at $q_c$, the barrier height being $E_b = E_c - Q_1$.

$E_c' = \frac{1}{2} M\omega_j^2(q_c-q_1)^2 + Q_1 = \frac{1}{2}M\omega_j^2 \frac{1}{4}(q_2-3q_1)(q_2+q_1)$ (28)

is the crossover energy between $V_1(q)$ and $V_2(q)$, while

$V_{12}(q_c) = b_{12}q_c = - (2M\omega_j^2)^{-1} b_{12}(b_1 + b_2)$ (29)

is the exchange term at crossover; use has been made of (12) in arriving at (27) from (28) and (29). Further, assuming hermiteicity of $V_{12}$ the adiabatic eigenstates (18) pertaining to the upper $E_U(q)$ and lower $E_L(q)$ surfaces have been obtained in the form [13]:

$| j,q >_U = \cos(\phi/2) | j_1,0 > + \sin(\phi/2) | j_2,0 >$

$| j,q >_L = -\sin(\phi/2) | j_1,0 > + \cos(\phi/2) | j_2,0 >$ (30)

where

$\sin\phi(q) = 2V_{12}(q) / R(q)$ and $\cos\phi(q) = [V_{11}(q) - V_{22}(q)] / R(q)$ (31)

with

$R(q) = ([V_{11}(q) - V_{22}(q)]^2 + 4|V_{12}|^2)^{1/2}$

standing for the square root in (20) and (21). Obviously, $\phi(0) = 0$ and $\phi(q_c) = \pi/2$.

Anticipating forthcoming needs, we next derive the curvature of $E_L(q)$ at $q_c$. It reads

$$d^2E_L/dq^2 = -M\omega_j^2[1+2(E_c-Q_1)/|V_{12}|] = -M\omega_j^2(b_2-b_1)^2/4|b_{12}||b_1+b_2| \quad (32)$$

use being made of (14), (28), and (22) on deriving (32) from (20). In absolute value this differs from the curvature of $E_L(q)$ at $q_j$ which is $M\omega_j^2$. From (26) we also get

$$|d^2E_L/dq^2| \gg \tfrac{1}{2}M\omega_j^2|(b_2-b_1)/(b_2+b_1)| \quad (33)$$

Finally, we define the lattice-displacement energy between the reorientational sites 1 and 2 by means of

$$E_r = \tfrac{1}{2}\{{}_{q2}\!\int^{q1}<1,0|-\partial V_e/\partial q|1,0>dq + {}_{q1}\!\int^{q2}<2,0|-\partial V_e/\partial q|2,0>dq\} \quad (34)$$

This is the average energy gained by lattice, as two reorientational sites are created at 1 and 2 by virtue of the electron-phonon interaction. From (3), (10), and (12) we get

$$E_r = -\tfrac{1}{2}(b_2-b_1)(q_2-q_1) = \tfrac{1}{2}M\omega_j^2(q_2-q_1)^2 \quad (35)$$

Comparing with (14) $E_r$ is also the value of the diabatic potential of a given reorientational site at the equilibrium position of the neighboring site, relative to their common energy minimum at $Q_1 = Q_2$. Relative to this same reference level the crossover energy (28) is a quarter of the lattice displacement energy (35): $E_c' - Q_1 = E_b' = \tfrac{1}{4}E_r$.

Next to solving the electronic problem (19), the vibronic eigenvalue equation

$$[T_L + E_j(q)]|j;n> = E_{jn}|j;n> \quad (36)$$

has to be dealt with, to obtain an approximate solution

$$|j,n> = |j,q>|j;n> \quad (37)$$

of the overall Schrodinger equation of an isolated dipole within the adiabatic approximation. Here $E_j(q)$ is given by either (20) or (21), consequently the corresponding vibronic eigenstates $|j;n>$ pertain to the lower or upper adiabatic surfaces. The eigenvalues of (36), $E_{jn}$, give the total energy of the dipole within the adiabatic approximation. Because of (26) $|j;n>$ will differ from the harmonic-oscillator eigenstates in the transition region around $q_c$ only. Finding the eigenvalues and eigenstates of (36) will give the energy levels and, ultimately, the transition probabilities pertaining to the reorientational problem.

## 3. Rationale of the reaction rate method

While the reorientational motion of a dipole actually occurs upon a multiwell potential surface along the 'relaxation coordinate', a two-site based analysis is usually found applicable, leading to $\tau_{rel} = (gk_{12})^{-1}$ where $k_{12}$ is the two-well relaxation rate and $g$ is an effective number of equivalent wells which depends on the symmetry of the dipole and the lattice, as well as on the orientation of the external field. To avoid complications, the relaxation coordinate is assumed rectilinear which makes it unnecessary taking into account centrifugal effects arising from the curvature. In any event, rectilinearity may be expected to provide a good approximation in a two-site approach to the relaxation process.

The usual theoretical prediction for the relaxation rate in a double-well situation would be based on the multiphonon approach (MPA) [13]. Mathematically, MPA rests on the time-dependent perturbation theory which defines the rate by means of Fermi's golden rule

$$k_{12} = (2\pi/h)\sum_{n1,n2} F(E_{n1},T) |<j_2,n_2| H' |j_1,n_1>|^2 \delta(E_{n2}-E_{n1}) \qquad (38)$$

where the sum is over the final states which conserve energy relative to the corresponding initial states, weighted by means of the thermal occupation probabilities $F(E_{n1},T)$. $H'$ is the relaxation-driving part of the Hamiltonian composed of the nuclear kinetic-energy operator (the nonadiabaticity operator). A further step is Condon's approximation which permits factorization of the matrix element in (38) into electronic and nuclear components

$$M_{12} = <j_2,n_2| H' |j_1,n_1> = <j_2,q| H' |j_1,q><j_2; n_2|j_1; n_1> \qquad (39)$$

Inasmuch as the transition probability is proportional to $|M_{12}|^2$, Condon's assumption effects factorization of that probability into electronic and nuclear parts.

The electronic part

$$K_{12} = <j_2,q| H' |j_1,q>$$

is assumed small to legitimize the use of the perturbation method. This confines the theory to transitions in which the initial and final electronic states $j_1$ and $j_2$ are only weakly coupled. Note that because of $q_1 \neq q_2$ (as a result of the electron-phonon interaction) the nuclear counterpart can be nonvanishing even though $n_1 \neq n_2$.

The reaction-rate approach (RRA), on the other hand, is based on an occurrence-probability formulation that accounts for both classical and quantal effects [10,14]. Now the transition rate is given by [15]:

$$k_{12} = \kappa(T)(k_BT/h)(Z_1^{\#}/Z)\exp(-E_c/k_BT) \qquad (40)$$

where

$$\kappa(T) = \sum_n W(E_n)\exp(-\varepsilon_n/k_BT)\Delta(\varepsilon_n/k_BT) \qquad (41)$$

is a functional factor to the rate equation, presented otherwise in its conventional classical form. $\kappa$, therefore, accounts for both nonadiabaticity and quantal effects. $Z_1$ is the complete partition function of the initial state, assumed to be in thermal equilibrium, $Z_1^{\#}$ is the initial-state partition function of the 'nonreactive' (accepting) modes obtained by excluding the relaxation-promoting mode from the domain of all the modes entering into $Z_1$. $E_c$ is the barrier energy (27) between the two neighboring reorientational sites at 1 and 2 along the relaxation coordinate. The motion along that coordinate is quantized with n, $E_n$ being the energy; $\varepsilon_n = E_n - E_c$ is the excess energy relative to the barrier peak, while $\Delta\varepsilon_n = E_{n+1} - E_n$ is the separation between two subsequent energy levels. $k_B$, h and T have their usual meaning.

Further,

$W(E_n) = \sum_{n'} W_{n'}(E_n) F(E_n, T)$

is the total transition probability

$W_{n'}(E_n) = \sum_{n''} W_{n'n''}(E_n)$

at level $E_n$ weighted over the nonreactive (accepting) quantum modes n' of the initial state $j_1$; n" are the respective modes in the final state $j_2$. Calculating $W(E_n)$ thus requires thermal averaging over the initial modes n'. However, $W_{n'}$ is independent of n' when the relaxation coordinate q is dynamically separable from the domain of all the mode coordinates; now $W(E_n) = W_{n'}(E_n)$. Under the present assumptions such is the case of the model in Section 2. Now, $W(E_n)$ is simply the transition probability at $E_n$ and calculating the quantum correction $\kappa$ reduces to an one-dimensional problem which has been solved [12]. The transition probability has also been assumed to factorize (Condon's assumption)

$$W(E_n) = W_L(E_n) W_e(E_n) \tag{42}$$

where $W_L$ is the probability for lattice rearrangement, while $W_e$ is the probability for a change of the electronic state at the transition configuration $q_c$ between states $j_1$ and $j_2$, under the conditions of energy conservation. Transitions with $W_e = 1$ are called 'adiabatic', while those with $W_e < 1$ are 'notadiabatic'.

Next, making use of the assumed harmonicity of the lattice vibrations, the promoting-mode contribution, which can be factorized out of $Z_1$ under the separability proposition, gives

$Z_1^\#/Z_1 = 2\sinh(h\nu_j/2k_BT)$.

Inserting into (40) one obtains finally

$$k_{12}(T) = \nu_{eff}^b \exp(-E_b/k_BT) \tag{43}$$

The preexponential factor in (43)

$$\nu_{eff}^b = \kappa(2k_BT/h\nu_j)\sinh(h\nu_j/2k_BT)\nu_j \tag{44}$$

can numerically deviate considerably from the rotational frequency $\nu_j$. Whether $\nu_{eff}^b/\nu_j > 1$ or $< 1$ depends essentially on the values of the correction factor $\kappa$, which is a temperature-dependent quantity. It should be stressed that both $\kappa > 1$ and $< 1$ cases are conceivable.

In a purely classical form of the rate theory

$$W_L(\varepsilon_n) = 1 \ (n \geq n_c), = 0 \ (n < n_c) \tag{45}$$

Because of (42), $\kappa$, now the electron transmission coefficient, is given by

$$k = {_0}\!\int^{\infty} W_e(\varepsilon)\exp(-\varepsilon/k_BT)d(\varepsilon/k_BT) \qquad (46)$$

For adibatic electronic transitions $W_e = 1$ and (46) yields $\kappa = 1$. (46) now ultimately gives $v_{eff}^b = v_j$ at sufficiently high temperatures ($k_BT \gg hv_j$). However, $\kappa$ may be significantly lower than unity if the electron transfer is nonadiabatic ($W_e \ll 1$). As a matter of fact, the electron-transfer probability for overbarrier transitions is given by [12]:

$$W_e = 2[1 - \exp(2\pi\gamma)] / [2 - \exp(2\pi\gamma)] \qquad (47)$$

where

$$\gamma(\varepsilon) = (V_{12}^2 / 2hv_j)E_r^{-1/2}\varepsilon^{-1/2} \qquad (48)$$

For a nonadiabatic relaxation $W_e \ll 1$ obtains at $\gamma \ll 1$ which gives $W_e = 4\pi\gamma$.

Inserting into (46) yields [15]:

$$\kappa = 4(V_{12}^2/hv_j)E_r^{-1/2}(k_BT)^{-1/2}\pi {_0}\!\int^{\infty} dx \exp(-x^2)$$

$$= 2(V_{12}^2 / hv_j)(\pi^3/E_r k_BT)^{1/2} \text{ for } V_{12} \ll E_b \qquad (49)$$

Comparing with (48) this gives $k = 4\pi^{3/2}\gamma(k_BT) = \pi^{1/2}W_e(k_BT)$. The electron-transmission coefficient is also predicted to decrease as the lattice-displacement energy $E_r$ increases. At $\kappa \ll 1$ the preexponential factor (44) will be much lower than the rotational frequency $v_j$. Consequently, too small preexponential factors observed experimentally could imply nonadiabaticity from the classical point of view.

From the quantum-mechanical point of view, however, the ratio $v_{eff}^b/v_j$ turns out to be rather different. Now, (49) gives but a lower limit to the nonadiabatic rate. At sufficiently high a temperature the tunneling subbarrier transitions ($\varepsilon_n < 0$) are as effective as the overbarrier classical jumps ($\varepsilon_n > 0$). This occurs in an energy range $E_n > E_c - k_BT_c$, where $T_c$ is 'Christov's characteristic temperature', defined by [12]:

$$T_c = hv_j^{\#}/\pi k_B \qquad (50)$$

$v_j^{\#}$ is an effective frequency given by the curvature at the barrier peak:

$$hv_j^{\#} = -\partial^2 E_L/\partial x_j^2 \qquad (51)$$

where

$$x_j = (M\omega_j^2/h\omega_j)^{1/2}q \qquad (52)$$

From (32) we now get

$$hv_j^{\#} = hv_j[1 + 2(E_c - Q_1) / |V_{12}|] = hv_j(b_2 - b_1)^2 / 4|b_{12}|b_1 + b_2| \qquad (53)$$

Because of (33) $\nu_j^\#$ generally exceeds $\nu_j$, $\nu_j^\# = \nu_j$ giving just a lower limit to the calculated characteristic temperature (50). Now, the following approximate expression has been obtained quasiclassically in the adiabatic limit for the quantum correction factor [12]

$$\kappa(T) = (\pi/2)(T_c/T)/\sin([\pi/2][T_c/T]) \text{ for } T > \tfrac{1}{2}T_c \tag{54}$$

Physically, this is a lattice-tunneling correction because the electron transfer occurs with certainty. Equation (54) shows $\kappa$ to be decreasing as T increases attaining values as low as 1 at large $T \gg T_c$. (Numerically $\kappa(T_c) = \pi/2$). This enables one to define the following temperature ranges characteristic of an adiabatic process:

(i) Quantal ($T < \tfrac{1}{2}T_c$) - tunneling-correction factor large;

(ii) Intermediate ($\tfrac{1}{2}T_c < T < 2T_c$) - tunneling factor close to 1 though still larger than 1;

(iii) Classical ($2T_c < T$) - tunneling factor equal to 1.

Accordingly, the dipolar relaxation occurs predominantly through subbarrier tunneling in (i), through nearly equal-weight tunneling and classical jumps (else, thermally-activated tunneling) in (ii), and via classical jumps in (iii).

Next, we use (43) and (54) to compute the activation energy

$$E_a = -\partial\ln\kappa_{12}/\partial(1/k_BT) \tag{55}$$

in range (ii). It is

$$E_a = E_b - (\tfrac{1}{2}h\nu_j)\coth(\tfrac{1}{2}h\nu_j/k_BT) + (\tfrac{1}{2}h\nu_j^\#)\cot(\tfrac{1}{2}h\nu_j^\#/k_BT) \tag{56}$$

For an estimate we set $T = T_c$ and $\nu_j^\# > \nu_j$, which is the usual occurrence. This gives

$$E_a = E_b - k_BT_c \tag{57}$$

The Arrhenius plot of an experimental dependence in the intermediate range will very nearly be a straight line, even though its slope will not yield the barrier height $E_b$ directly. The latter obtains from the plot in the classical range for an adiabatic process, as it follows from (56) at $T \gg T_c$. To compute the preexponential factor

$$\nu_{eff}^a = k_{12}\exp(E_a/k_BT) \tag{58}$$

in the intermediate range we use (57) to obtain approximately at $T \sim T_c$,

$$\nu_{eff}^a = \kappa(T)(2k_BT/h\nu_j)\sinh(\tfrac{1}{2}h\nu_j/k_BT)\nu_j\exp(-T_c/T) \tag{59}$$

Equations corresponding to (57) and (59) for $\nu_j^\# = \nu_j$ obtain from (56) and (58) likewise.

## 4. Reassessment of experimental data from the RRA viewpoint

The remainder of this paper shall deal with a reassessment of some experimental data from the viewpoint of the theory outlined in the preceding Sections. We will concentrate on data obtained by the ITC method. Physically it consists of polarizing the dipolar system in an external electric field $E_p$ at some temperature $T_p$ until a saturated polarization $P_s = Nd p^2 E_p /3k_B T$ is achieved [11]. Following this the sample is cooled down to a temperature $T_0 \ll T_p$, where the dipolar polarization P is frozen-in, and the electric field switched off. Now, the system is gradually warmed up at a constant rate $\beta = dT/dt$ and the depolarization current $J = dP/dt$ measured as a function of time t (temperature T). If $\tau$ is the relaxation time for dipolar reorientation, the polarization varies in time according to

$$dP/dt + P/\tau = 0 \quad \text{or} \quad P(t) = P_s \exp(-\int_{t_0}^{t} dt/\tau) \tag{60}$$

It should be noted that since $\tau$ is a temperature- and, therefore, time-dependent quantity, it is an essential and generally unknown part of the integrand in (60). The problem further is to express $\tau(t)$ in terms of experimentally measurable quantities without making any specific assumptions with regard to the particular temperature dependence involved. Inasmuch as from (60) $J(t) = -P/\tau$ we get

$$\int_t^\infty J(t)dt = P(\infty) - P(t) = \tau(t)J(t)$$

assuming $P(\infty) = 0$. Consequently,

$$\tau(T) = (1/\beta) \int_T^\infty J(T')dT' / J(T) \tag{61}$$

as a function of the temperature $T = T_0 + \beta(t-t_0)$. Equation (61) forms the basis of the experimental determination of the relaxation time through portional integration of an ITC band.

The additional assumption $\tau = \tau_0 \exp(E_b/kT)$ at constant $\tau_0$ is often made as far as a classical reorientational behavior is expected which helps to simplify the experimental procedure through evaluating $E_b$ and $\tau_0$ from the peak temperature $T_m$ and halfwidth W of the ITC band. In the light of the theory presented in Section 3, the above assumption should be regarded as misleading and must therefore be avoided. Fortunately, some experimentalists have correctly preferred measuring the entire temperature dependence, making use of (61), which does not mask the quantal effects.

A summary of experimental data on the relaxation of I-V dipoles in $Eu^{2+}$-doped alkali halides obtained at these Laboratories are presented in Table I [16]. The data are arranged in an increasing order of the activation energies (and of the pre-exponential factors, as well), both calculated from the ITC band parameters assuming a classical rate process. Also listed are the interionic separations and the LO-phonon frequencies of the corresponding host materials. The $\nu_{LO}$'s are seen to generally exceed the respective pre-exponential factors $\nu_{eff}^a$ in a decreasing trend from the top down to the KCl dipole where the trend is reversed ($\nu_{LO} < \nu_{eff}^a$). Due to the lack of more specific data on the rotational frequencies, we shall further base our analysis on the LO-data assuming $\nu_j = \nu_{LO}$. Unlike the findings for other impurity-host systems exhibiting a rather complicated reorientational path [18], a single rotation mechanism will be assumed for the $Eu^{2+}$-$V_c$ dipole providing for a direct application of the theory outlined in the preceding Section.

We take up the case of the first seven dipoles in Table I initially. The small $\nu_{eff}^a / \nu_{LO}$ ratios suggest a nonadiabatic behavior. So strong it seems to be, that any hidden quantal effects are eventually masked, especially in the uppermost cases, even though they have exhibited low apparent barrier heights. To simplify the matter, we shall presume classical nonadiabaticity throughout. Now, $E_a = E_b$, while the electron-transmission coefficient $\kappa$ is given by eq. (49). To calculate $\kappa$ at the peak temperature $T_m$ in each case, we make use of eq. (44). This results in the second line of Table II. Next, we assume $E_r = 4E_b' = 4E_b$ to fill in the third line, using the data of the former line, the fourth column of Table I, as well as eq. (49). The corresponding deviations from nonadiabaticity at the peak temperatures $T_m$ are calculated as $1 - W_e(k_B T_m)/4\pi\gamma(k_B T_m) = 1 - W_e(k_B T_m)/\pi^{-1/2}\kappa(k_B T_m)$, where $W_e$ is given by eq.(47) at $\gamma = (4\pi^{3/2})^{-1}\kappa$ and listed in the fourth line. The dipoles are seen to become increasingly less nonadiabatic as the apparent barrier height increases from left to right. No immediated relationship can be established with the variations of the interionic separation in that same order unlike the reorientation of the $OH^-$ quantum dipole where a $V_{12} \sim \exp(-\text{const} \times \text{interionic spacing})$ dependence has been observed [19].

The reason for the lack of any correlation with the lattice parameter may partially be in that the contribution of the lattice tunneling may have been increasingly underestimated, as one goes from left to right in Table III. This is clearly manifested by the last (KCl) line in Table I. Now the large $\nu_{eff}/\nu_{LO}$ suggests an intermediate-dipole occurrence. We presume adiabaticity ($W_e = 1$) for the KCl dipole and use relevant equations of Section 3 to calculate the pertinent parameters. In so far as the apparent barrier height $E_a$ in the intermediate range is given by eq.(57), at least about the characteristic temperature, we rewrite

$$k_a(T) = \exp(-T_c/T)(\pi/2)(T_c/T)/\sin([\pi/2][T_c/T]) \qquad (62)$$

before inserting into (59) and use the experimental data in Table I to obtain $T_c$. Once this has been done, we make use of (57) to calculate the actual barrier height $E_b$, and thereby the electron-transfer term $|V_{12}|$ from

$$T_c = (\nu_{LO}/\pi k_B)(1 + 2E_b/|V_{12}|) \qquad (63)$$

which derives from (50) and (53). Calculating $E_b' = E_b + |V_{12}|$ and $E_r = 4E_b'$ then follows straightforwardly. The results of all these calculations are presented in Table III. From the obtained lattice-displacement energy $E_r$ one can compute using (35) the magnitude of the displacement between the well minima, in terms of the dimensionless lattice coordinate $\xi$ from (52), since the force constant of the promoting-mode oscillator is unknown. Nevertheless, one arrives at $q_2 - q_1 = 2$ Å on using the LO data (3.12 eV/A$^2$) [20], not too bad in view of the approximations involved. Finally, using the parameters presented in Table III, one can construct the configurational-coordinate (CC) diagram pertinent to the reorientation of the I-V dipole in KCl:Eu$^{2+}$. The result is shown in Fig.1 plotting both the lower and the upper adiabatic branches versus the dimensionless phonon coordinate (52). A similar CC-diagram of a nonadiabatic classical I-V dipole in KI:Eu$^{2+}$, calculated using the Table II data (fourth column) is presented next in Fig.2.

The entire temperature dependence of the relaxation time of the KCl dipole is reconstructed, using equations (43) and (54) and the second-line data in Table III, as shown in Fig.3. The sole circle therein represents the starting point of the calculation, as given by $\tau(T_m) = \nu_{eff}^{a-1} \exp(E_a/k_B T_m)$ and the Table I data. The theoretical curve is seen to pass some 25% higher than the circle at $T_m$ due to the correction introduced leading to the actual barrier height $E_b$, in accordance with equation (57). The predicted quantal range of reorientation is seen to extend below 140 K and involve relaxation times in excess of $10^9$ s.

## 5. Comment on the photostimulated dipolar relaxation

A number of I-V dipolar systems have exhibited a remarkable enhancement of the reorientational rate upon photoexcitation in the intrinsic absorption bands of the impurity. Examples of such a photostimulated dipolar relaxation have experimentally been found in both the cation and anion sublattices [21,22]. Speculations have been made on the physical background of the observed rate enhancement pointing to either a change of barrier height $E_b$ or preexponential factor $v_{eff}^b$ or both [22,23]. The only quantitative data that are available reveal a drop of barrier from 0.7 eV down to 0.1 eV on photoexcitation of the Suzuki-phase bound-dipole system in NaCl:$Pb^{2+}$ [21]. This does not seem to be any peculiar property of Suzuki phases only.

From a RRA viewpoint both $E_b$ and $v_{eff}^b$ can change on passing from an electronic ground state j = g to an excited state j = e. Ultimately, both changes would result from corresponding variations of the electron-phonon coupling coefficients, as well as of the electron-exchange term. To focus our attention on $E_b$ first, we reproduce the resulting equation in either state 'j':

$$E_b = (|b_2 - b_1| / 2M\omega_j^2)\{¼|b_2 - b_1| - |b_{12}||(b_2 + b_1)/(b_2 - b_1)|\} \qquad (64)$$

$$= (|b_2|^2 / 2M\omega_j^2)\{¼ - |b_{12}| / |b_2|\} \text{ for } b_1 = 0$$

Because of (26), (64) defines a positive quantity as it should. In principle, a drop of barrier as one goes from 'g' to 'e' could therefore result from:

(i) an increase of the dressed-socillator frequency $\omega_j$ brought about by a corresponding change of the second-order coupling coefficient $c_j$ (an increase for $c_j > 0$ or a decrease for $c_j < 0$);

(ii) a decrease of the linear-coupling coefficient $|b_j|$;

(ii) an increase of the electron-exchange term $|b_{12}|$.

As a matter of fact, an increase of the coupled-oscillator frequency $v_j$ is often observed on going to the e.e.s. of the component defect [17]. A more rigorous change of $|b_j|$ and $|b_{12}|$ according to (ii) and (iii) would bring about a drop in characteristic temperature, following (53), that is,

$$T_c = ¼(hv_j / \pi k_B)(b_2 - b_1)^2/|b_{12}||b_1 + b_2| \qquad (65)$$

$$= ¼(hv_j / \pi k_B)(|b_2| / |b_{12}|) \text{ for } b_1 = 0$$

with the result of an adiabatic dipole being eventually moved to the classical range thereby enhancing its relaxation rate. It is to be stressed that under the above conditions ((i) to (iii)) the lattice-displacement energy $E_r$ drops too following (35):

$$E_r = ½(M\omega_j^2)^{-1}(b_2 - b_1)^2 \qquad (66)$$

reflecting the fact that the two diabatic parabolae (14) at '1' and '2' displace towards each other more effectively than they get steeper in the e.e.s.

Although the behavior of the coupling constant and the exchange term can only be quantitatively assessed based on a proper vibronic model, we are tempting to speculate, nevertheless, that as the electronic state of the impurity gets more extended, coupling to the promoting mode weakens in magnitude. This implies that $c_j < 0$, while the local mode itself involves the vibrational motion of e.g. the <110> nn-cation if driving a dipole of <110> symmetry. Such a general conclusion is supported by some calculations on the breathing-mode coupled F center in alkali halides [24]. At the same time, an increase of $|b_{12}|$ may result from the increased wave function overlap due to the larger extension of the electron clouds in the e.e.s.

## 6. Discussion

The purpose of the present paper has neither been to give a comprehensive expose of the reaction-rate method, not to define precisely its applicability limits to dipolar-relaxational problems. While literature on the former has long been available [10,12,15], the latter is still under study, and is perhaps not immune against misinterpretations which come naturally each time a theory is originally transferred from a different, though sister branch of science - chemistry, to a field so inherently belonging to physics. Some first steps in this direction have already been taken definitely [8,9,19], and the results obtained have proved encouraging enough to stimulate further step-by-step attempts. It is in this context that the present work on the IV-dipoles should be regarded and judged: It has aimed at digging some novel information out of a field where classical interpretations have traditionally been so firmly rooted.

To begin with the Hamiltonian background in Section 2, we have first defined a model energy-operator of an isolated dipole that is simple enough so as not to obscur physics. Apart from the approximations mentioned therein, the more realistic angular relaxation coordinate has been replaced by a rectilinear one, to avoid complications arising from the curvature. However simplified, it may be expected to work pretty well in a two-well analysis of relaxation-time data. Next we have preferred to follow a way of solving Schrodinger's equation in the adiabatic approximation through first defining the diabatic potentials by means of the static electronic eigenstates $|j,0>$, unlike the procedure used by Christov [12] and then introducing an adibatic eigenstate $|j,q>$ as a linear combination of $|j1,0>$ and $|j2,0>$, corresponding to the two neighboring reorientational sites '1' and '2' [25]. $|j1,0>$ and $|j2,0>$ themselves, may be assumed to form e.g. as symmetric and antisymmetric combinations of Slater's orbitals, respectively, as proposed at its time by Luty [6], arise self-consistently because of the tunneling interaction between '1' and '2' and play the essential role of defining these sites. The above procedure, similar to the one proposed by Pässler [13], has enabled us to define the electron-exchange term V12 by means of the off-diagonal matrix element of the first-order electron-lattice coupling operator, thereby avoiding the necessity of using a donor-acceptor model [10], perhaps less immediately applicable to the dipolar case. Somewhat differing from Pässler's approach, we have introduced an electron-lattice coupling Hamiltonian (4) (dots omitted) in a up-to-the-second-order form, to allow 'dressing' of the promoting-mode oscillator through its coupling to the electronic state 'j', thereby accounting for possible changes of the vibrational frequency (13) as that state changes. Of course, this has directly been intended for an application to the photostimulated relaxation, touched only briefly, after all, because of the scarcity of available experimental data, and the lack of any reliable vibronic model at the time. However, in defining the exchange term $V_{12}$ we have again adopted Passler's linear-coupling scheme, confining in effect the expansion (22) to the first-order term [26]. Finally, to simplify the problem even further, we have presumed a symmetric-well situation, in which the reorientational hopping involves no zero-point reaction heat at all. This implies hopping of a dipole between sites of the same symmetry, and is believed to apply to the experimental examples dealt with in Section 4, although it certainly does not to more complex reorientational paths comprising both rotation and translation, as found for other dipolar systems. [5,18].

Turning next to the rationale of the method in Section 3, we have started by comparing the RRA rate equation (40) with the one that derives from the MPA perturbation method, to underline their basic difference: While MPA applies to transitions between weakly coupled electronic states, RRA works equally well for all coupling strengths. Care must, however, be taken in the quantitative appraisal of that statement, since the coupling operators are different in the two cases. In exposing the mathematical basis of the RRA formalism we have generally followed Christov's brilliant way [10,12,15]. Perhaps the only substantial step aside has been taken on defining the effective frequency $v_{eff}^{\#}$ by means of the actual curvature (32) at $q_c$ of the lower adiabatic surface (20). This has led to a realistic equation (63) for the characteristic temperature $T_c$. It should also be pointed out that eq. (59) for $v_{eff}^{a}$ holds true in the intermediate range about the characteristic temperature only, so that care should be taken on extending its applicability beyond. The substance covered under (i) to (iii) in Section 3 introduces an important distinction between the dipoles, based on their observed relaxational behavior, which helps to classify them according to the type of electron transfer and lattice rearrangement involved.

The only reason why we have reproduced the formulae leading to eq. (61) of Section 4 has been for the sake of showing that they do not at all necessitate presuming any classical temperature dependence for the relaxation time $\tau(T)$. Thus, eq. (61) can well be used for obtaining the complete accessible information on the relaxation path, whether classical, intermediate, or quantal. This has often been underestimated by experimentalists. Further, the data listed in Table I have been obtained assuming a classical relaxation rate [16]. The presumed classical dipoles in Table II are seen to exhibit a fair non-adiabaticity, which, however, gets increasingly worse as one goes from left to right. This has prompted the attempt to treat the KCl dipole (exhibiting a deviation from non-adiabaticity of as much as 76 percent, calculated as above assuming a classical occurrence) as an intermediate adiabatic entity. Using the relevant Section 3 formulae we have first calculated the characteristic temperature $T_c$ from the experimental value of $v_{eff}^{a}$ and eq. (59), where $\kappa(T)$ is given by eq. (54), while $E_a$ has been assumed to be the one from the fourth column of Table I. This procedure, correct only near the characteristic temperature $T_c$, has yielded a $T_c$ that is a little less than twice the ITC-peak temperature $T_m$. Clearly, though still in the intermediate range, this has introduced an error in the actual barrier height $E_b$, estimated by means of eq. (57), which amounts to $(E_b^{corr}-E_b)/E_b$ = 36%. If, however, $k_B T_c$ in eq. (57) is assumed to constitute just a small correction to $E_b$, and eq. (44) with $v_{eff}^{b} = v_{eff}^{a}$ ($E_b \approx E_a$) is used to calculate $T_c$ from (54), the result is $T_c$ = 276.37 K, $E_b$ = 0.68 eV (from (57)), and the error is less than 3%. Clearly, the accurate determination of $E_b$ should be subject of an iteration procedure involving eq's (58) and (63). It should be noted that the above analyses of the $Eu^{2+}$ I-V data have been based on the presumption that g = 1, i.e. $\tau_{rel} = k_{12}^{-1}$, as done elsewhere [27]. Thus the results obtained should only be regarded as rough estimates, let alone the actual rotational frequencies which are not precisely known.

Although the present single-frequency harmonic-oscillator model may be oversimplified, it displays all the basic features of the reaction-rate method. Accordingly, we hope to have demonstrated the RRA capacity of providing a useful means for interpreting observed relaxational behavior which often goes beyond, and contains more information than, the classical adiabaticity that is usually implied. Instead, in some cases complicated reorientational mechanisms have been invoked, rather than simply taking into account the influence of the upper adiabatic surface, to explain low pre-exponential factors measured experimentally [28]. In other cases, quantal or intermediate effects may have remained undisclosed because of data processing based on presumed classical conduct. To avoid it, use of eq. (61) for obtaining the relaxation time is strongly recommended.

Generally, the present dipole reorientation model contains a few features which may raise doubts if viewed as gross oversimplifications in the light of the lower symmetry of the species. One is the electron-mode coupling which regards the dipole as a single-atom entity rather. Clearly, this is permissible if the impurity-vacancy interchange rate is largely inferior to the vacancy intersite jump rate at frozen impurity. The single-atom approach has been applied to and found useful for describing the reorientation of off-center dipoles around their normal lattice sites in alkali halides [30]. Another one is the 1-D approach to a multimode problem though its deficiencies should not be overestimated too. Indeed, the reorientation of in-plane off-center dipoles in colored $KCl:Li^+$ has been found tractable by the transcendent Mathieu equation which is 1-D in the azimuthal angular coordinate [29].

Finally, ESR experiments on the temperature dependence of the spin Hamiltonian parameters lend support to our suggested <110> symmetry of the $Eu^{2+}$ dipole in KCl [30].

References

[1] V. Narayansmurti and R.O.Pohl, Revs. Mod. Phys. **42**, 201 (1970).
[2] A.S. Barker and A.J. Sievers, Revs. Mod. Phys. **47**, S1 (1975).
[3] O. Kanert, Phys. Reports **91**, 183 (1982).
[4] F. Luty, J. Physique (Paris) **28**, C4-120 (1967); **34**, C9-49 (1973).
[5] A.S. Nowick, in Point Defects in Solids, edited by J.H. Crawford, Jr. and L.M. Slifkin (Plenum, New York-London, 1972), p. 151.
[6] F. Luty, in Physics of Color Centers, edited by W. Beall Fowler (Academic, New York, 1968), p. 181.
[7] See H.B. Shore and L.M. Sander, Phys. Rev. B **6**, 1551 (1972).
[8] A. Gochev, Solid State Commun. **49**, 1181 (1984).
[9] M. Georgiev, J. Signal AM (1984) (accepted).
[10] S.G. Christov, Collision Theory and Statistical Theory of Chemical Reactions (Springer, Berlin-Heidelberg-New York, 1980)
[11] R. Capelletti and R. Fieschi, Crystal Lattice Defects **1**, 69 (1969).
[12] S.G. Christov, Phys. Rev. B **26**, 6918 (1982).
[13] R. Passler, Czech. J. Phys. B **32**, 846 (1982).
[14] R. Holstein, Phil. Mag. **37**, 49; 499 (1978).
[15] S.G. Christov, Berichte Bunsenges. Phys. Chem. **79**, 357 (1975).
[16] J. Hernandez A., H. Murrieta S., F. Jaque, and J. Rubio O., Solid State Commun. **39**, 1061 (1981).
[17] R.S. Knox and K.J. Teegarden, in Physics of Color Centers, ed. by W. Beall Fowler (Academic, New York, 1968), p. 625; W. Beall Fowler, ibid., p. 627.
[18] S. Radhakrishna and S. Harridos, Crystal Lattice Defects **7**, 191 (1978).
[19] M. Georgiev and A. Gochev, Phys. Rev. B (1984) (accepted).
[20] M. Georgiev, A. Gochev, S.G. Christov, and A. Kyuldjiev, Phys. Rev. B **26**, 6936 (1982).
[21] R. Capelletti, in Transport Properties of Solids, edited by E. Mariani (Institute of Physics, Bratislava, 1982), p. 15.
[22] J. Prakash, Phys. Status Solidi B **114**, 61 (1982).
[23] F. Fisher, Phys. Status Solidi A **52**, 189 (1979).
[24] M. Georgiev, Phys. Rev. B (1984) (accepted), and unpublished.
[25] An alternative way would be to introduce the electron-exchange operator in the basic Hamiltonian, as done by Holstein [14]. We chose the present sequence for 'historical reasons' mainly.
[26] Taking into account some higher-order terms would undoubtedly open new possibilities that have not yet been explored.
[27] M.S. Li, M. de Souza, and F. Luty, Phys. Rev. B **7**, 4677 (1973).
[28] C. Bucci, Phys. Rev. **164**, 1200 (1967).
[29] P. Petrova, M. Ivanovich, M. Georgiev, M. Mladenova, G. Baldacchini, R.M. Montereali, U.M. Grassano, A. Scacco in: *Quantum Systems in Chemistry and Physics*, R. McWeeny et al., eds. (Kluwer Academic Publishers, Dordrecht, The Netherlands), pp. 373-395 (1997).
[30] C. Pina, E. Munoz, and J.L. Boldu, J. Chem. Phys. **79**, 2172 (1983).

TABLE I. Summary of experimental data on the reorientation of $Eu^{2+}$ I-V dipoles in alkali halides, after Refs. 16 & 17. The data are arranged in the increasing order of the apparent barrier height $E_a$

| Host crystal | Interionic Separation d (Å) | ITC Peak Position $T_m$ (K) | Apparent barrier $E_a$ (eV) | Pre-*exp* factor $\nu_{eff}$ ($\times 10^{12}$ s$^{-1}$) | LO-phonon frequency $\nu_{LO}$ ($\times 10^{12}$s$^{-1}$) | LO-phonon frequency $\omega_{LO}$ ($\times 10^{13}$s$^{-1}$) |
|---|---|---|---|---|---|---|
| NaI | 3.237 | 177 | 0.18 | $1.6\times10^{-9}$ | 5.52 | 3.47 |
| NaBr | 2.989 | 188 | 0.33 | $1.2\times10^{-5}$ | 6.33 | 3.98 |
| RbBr | 3.445 | 195 | 0.39 | $3.3\times10^{-4}$ | 3.90 | 2.45 |
| KI | 3.533 | 197 | 0.46 | $1\times10^{-2}$ | 4.30 | 2.70 |
| RbCl | 3.291 | 198 | 0.49 | $1\times10^{-1}$ | 5.41 | 3.4 |
| KBr | 3.298 | 210 | 0.54 | $2\times10^{-1}$ | 5.11 | 3.21 |
| NaCl | 2.820 | 217 | 0.58 | $5\times10^{-1}$ | 8.12 | 5.1 |
| KCl | 3.147 | 219 | 0.66 | $1.5\times10^{+1}$ | 6.40 | 4.02 |

TABLE II. Calculated RRA characteristics of presumed nonadiabatic classical $Eu^{2+}$ I-V dipoles in alkali halides using the data listed in Table I. See text for further details.

| Host | NaI | NaBr | RbBr | KI | RbCl | KBr | NaCl |
|---|---|---|---|---|---|---|---|
| $\kappa(T_m)$ | $2.6\times10^{-10}$ | $1.7\times10^{-6}$ | $8.1\times10^{-5}$ | $2.2\times10^{-3}$ | $1.7\times10^{-2}$ | $3.7\times10^{-2}$ | $5.4\times10^{-2}$ |
| $|V_{12}|$(eV) | $2.8\times10^{-7}$ | $2.1\times10^{-5}$ | $1.1\times10^{-4}$ | $5.8\times10^{-4}$ | $1.8\times10^{-3}$ | $2.5\times10^{-3}$ | $3.8\times10^{-3}$ |
| $\Delta A$ (%) | < 1 | < 1 | < 1 | < 1 | 2 | 5 | 7 |
| $E_b'$ (eV) | 0.18 | 0.33 | 0.39 | 0.46 | 0.49 | 0.54 | 0.59 |
| $E_r$ (eV) | 0.72 | 1.32 | 1.56 | 1.84 | 1.97 | 2.18 | 2.35 |
| $\sqrt{(2E_r/h\nu_{LO})}$ | 7.94 | 10.04 | 13.91 | 14.41 | 13.28 | 14.36 | 11.82 |

$\kappa(T_m)$ - rate correction; $|V_{12}|$ - crossover energy gap; $\Delta A$ - deviation from nonadiabaticity
$E_b'$ - crossover energy; $E_r$ – lattice displacement energy; $\sqrt{(2E_r/h\nu_{LO})}$ – lattice displacement

TABLE III. Calculated RRA characteristics of the presumed adiabatic intermediate $Eu^{2+}$ I-V dipole in KCl using the data of the last line in Table I. See text for details.

| $h\nu_{LO}/\pi k_B$ (K) | $\kappa_a(T_m)$ | $T_c$ (K) | $E_b$ (eV) | $|V_{12}|$ (eV) | $E_b'$ (eV) | $E_r$ (eV) | $\sqrt{(2E_r/h\nu_{LO})}$ |
|---|---|---|---|---|---|---|---|
| 97.74 | 2.16 | 408.54 | .69±36% | 0.44 | 1.13 | 4.53 | 18.50 |
|  |  | 276.37 | .68±3% | 0.75 | 1.43 | 5.72 | 20.80 |

The upper and lower sets in columns 3 through 8 arise on using equations (59) and (44), respectively, to calculate Christov's characteristic temperature $T_c$ from the experimentally observed pre-exponential factor in column 5 of Table I. In both cases $\kappa(T_m)$ has been given by eq. (54) at the ITC peak temperature $T_m$. The actual barrier heights $E_b$ have been calculated using eq. (57) and the experimental value of $E_a$ from column 4 of Table I.

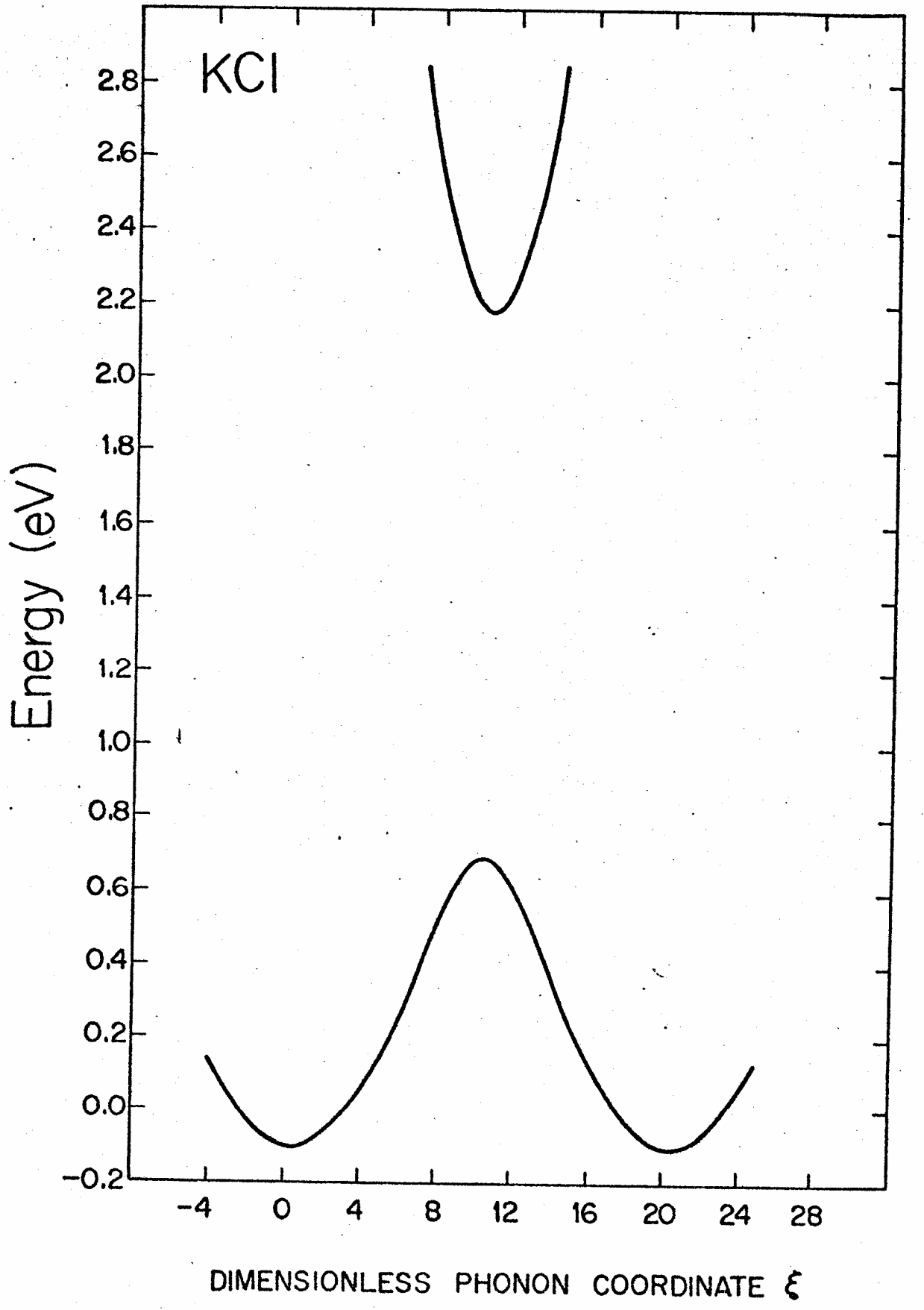

Figure 1

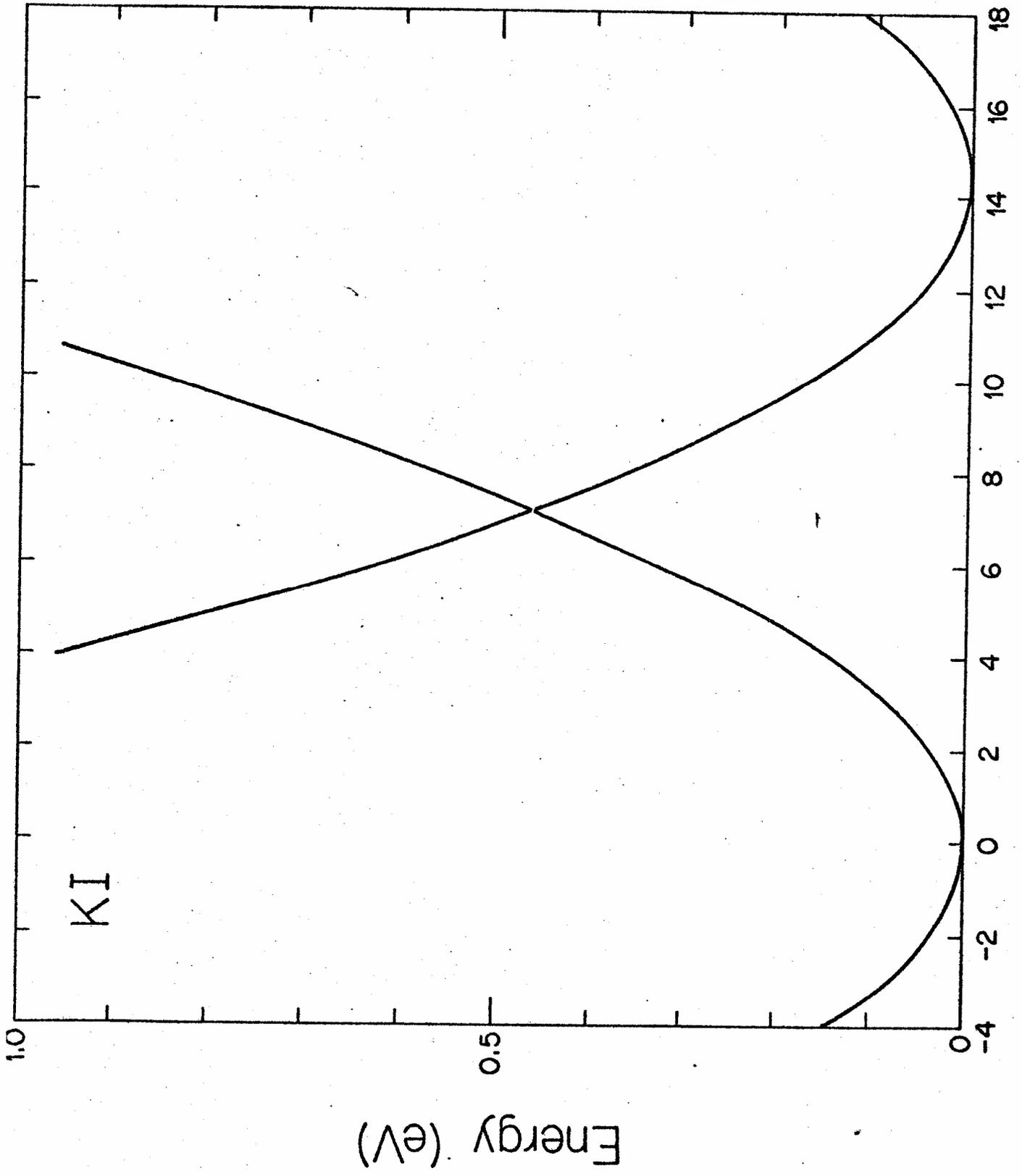

Figure 2

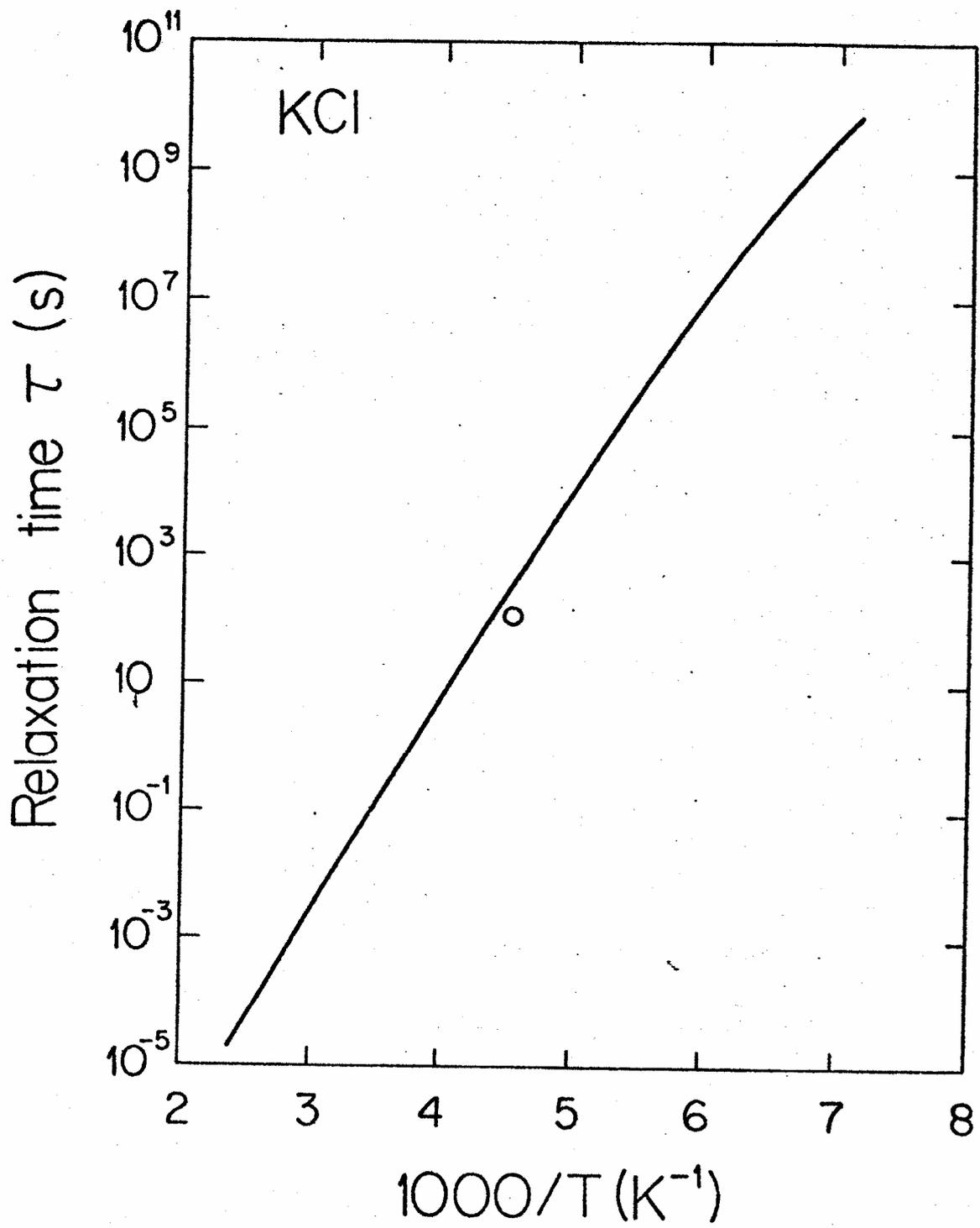

Figure 3

Figure captions

Figure1. Adiabatic potentials pertinent to the reorientation of adiabatic intermediate dipole in KCl:Eu$^{2+}$ versus the dimensionless phonon coordinate $\xi$, as calculated using the second set of data in Table III. The equilibrium coordinate of the left-hand well is assumed to be at $\xi_1 = 0$ ($b_1 = 0$). The gap between the upper and the lower potential energy branches at saddle point amounts to twice the electron-exchange term $|V_{12}|$. Inasmuch as the latter is large, it is the saddle-point barrier between reorientational sites that limits the relaxation rate.

Figure 2. Adiabatic potentials pertinent to the reorientation of nonadiabatic classical dipole in KI:Eu$^{2+}$ versus the dimensionless phonon coordinate $\xi$ calculated using the fourth-column data of Table II. The equilibrium coordinate of the left-hand well is set at $\xi_1 = 0$ ($b_1 = 0$). The gap between upper and lower potential energy branches at saddle point amounts to twice the electron-exchange term $|V_{12}|$. The latter being small, the proximity of the upper potential surface now greatly reduces the relaxation rate, even though the barrier height is lower.

Figure 3. Predicted temperature dependence of the relaxation time of the I-V dipole in KCl:Eu$^{2+}$ calculated using eq's (43) and (54), and the second-line data in Table III. The sole circle represents the experimental relaxation time extracted from the ITC peak temperature and halfwidth assuming a classical reorientational process (Ref. 16).